\documentclass[final,12pt]{elsarticle}

\usepackage{amssymb,amsmath}

\usepackage{color}
\usepackage{lineno}
\usepackage{eurosym}
\usepackage{hyperref}
\journal{Journal of Optics}

\begin{document}

\begin{frontmatter}

\title{Single Pixel Imaging and Compressive Sensing: A Practical Tutorial}

\author[inst1]{Dennis Scheidt}\ead{d.scheidt@fz-juelich.de}

\affiliation[inst1]{organization={Institute for Neuroscience and Medicine 1, Forschungszentrum Juelich},
            addressline={Wilhelm-Johnen-Straße}, 
            city={Juelich},
            postcode={52428}, 
            state={NRW},
            country={Germany}}

\begin{abstract}
\noindent 
Single Pixel Imaging is an emerging imaging technique that employs a bucket detector (photodiode) to sample a spatially modulated light field, rather than measuring the spatial distribution with an array of detectors. This approach provides a low-cost alternative for imaging at unconventional wavelengths and enables improved signal collection in noisy measurement environments. Furthermore, it allows the application of compressive sensing to reduce the amount of acquired data and measurement time, facilitating live or \textit{in vivo} imaging applications. This tutorial presents the experimental implementation of measurement bases and compressive sensing reconstruction methods, including both deterministic algorithms and deep learning approaches. Accompanying Python notebooks guide readers through the reproduction of the presented results and support the application of the methods to their own work.
\end{abstract}


\end{frontmatter}


\section{Introduction}

\noindent
Single Pixel Imaging (SPI) was introduced in 2006 by Baraniuk and co-workers \cite{Duarte} as a method to reconstruct images using a single photodetector, in contrast to conventional cameras based on CCD or CMOS sensor arrays. In SPI, an object is sequentially illuminated with spatial patterns, typically defined by an orthogonal basis, using a digital micromirror device (DMD). The light reflected or transmitted by the object is collected and measured by a single photodiode (bucket detector). By recording the intensity associated with each pattern, the image can be reconstructed. A key advantage of SPI is its applicability at wavelengths where CCD or CMOS sensors are inefficient or unavailable, providing a cost-effective alternative for such spectral regions.

\noindent
The conceptual foundations of SPI can be traced back to ghost imaging, developed in the 1990s by Klyshko and Shih \cite{PhysRevLett.74.3600,PhysRevA.52.R3429}, where spatial information is retrieved from intensity correlations. Building on this idea, Computational Ghost Imaging (CGI) replaces correlated photon pairs with spatial light modulators (SLMs) that encode known patterns onto the illumination field \cite{Padgett_CGI}. CGI has been extended to holographic techniques enabling amplitude and phase measurements \cite{PhysRevA.86.041803}. Although SPI and CGI differ primarily in their illumination schemes, they are closely related and are treated as equivalent for the purposes of this tutorial \cite{Padgett_CGI,GONG2022108140}. Similar single-pixel approaches have also been combined with interferometry to measure the phase of optical fields, where intensities are recorded at a single camera pixel or through spatial filtering followed by photodetection \cite{SinglePixel,Ota:18,Zupancic:16,Sephton:23_GhostImagingPhase}.

\noindent
A major advancement in SPI is the incorporation of compressive sensing (CS), which enables image reconstruction from a subset of the measurement basis elements by exploiting signal sparsity. CS-based SPI has been successfully applied to complex wavefront imaging \cite{Clemente:13,Horisaki:17_2,Howland:14} and to enhance light transmission through scattering media \cite{T_Mat3,Tajahuerce:14}. The high refresh rates of modern DMDs (several kHz) have further enabled applications such as dynamic wavefront correction \cite{SinglePixel}, retinal imaging \cite{Dutta:19}, and light focusing through biological tissue and random scattering media \cite{Wang:Mouse,Boniface:19}. Recent work has also demonstrated that the ordering of Hadamard bases can significantly affect reconstruction quality in compressive SPI \cite{HadOrder,HadRussDoll,HadOrigami}.

\noindent
Conventional CS reconstruction typically relies on $\ell_1$-minimization, which can be computationally expensive and often slower than the data acquisition itself \cite{databook,Yang_L1mini}. As an alternative, deep learning approaches shift the computational burden to the training phase, after which reconstructions can be performed in milliseconds on standard CPUs \cite{YAO2019483}. This enables high-speed and real-time SPI applications, including kHz-rate image reconstruction \cite{Wang:Mouse,SPI_HadFormula,Boniface:19}.

\noindent
This tutorial first describes the experimental implementation of SPI using digital micromirror devices (DMDs) and spatial light modulators (SLMs) (Sec.~\ref{sec:Experiment}). It then addresses image reconstruction using different measurement bases, employing both deterministic compressive sensing algorithms (Sec.~\ref{sec:CS}) and deep learning methods (Sec.~\ref{sec:DL}). Two Python notebooks are provided to reproduce the \href{https://colab.research.google.com/drive/1BOQHEbJbRKe9fbTBnLtMMlTHFeOjXtF8?usp=sharing}{reconstruction} and \href{https://colab.research.google.com/drive/1d0hrh3v9VdLLxFooUzy88cT48RY8_TWf?usp=sharing}{training} procedures presented in this tutorial Python notebooks.

\section{Methods}\label{sec:Experiment}

\subsection{Experimental Setup}

\begin{figure}[h]
    \centering
    \includegraphics[width=\linewidth]{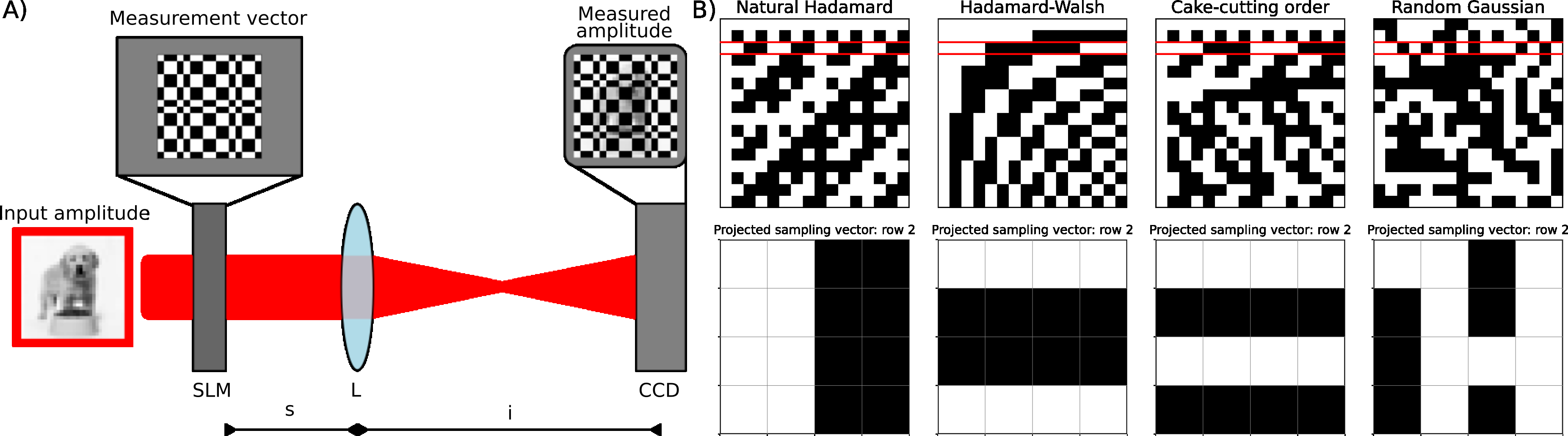}
    \caption{\textbf{A)} Basic setup for Single Pixel Imaging.
    The input field $x$ is represented by the dog picture. On the SLM a row vector of the Hadamard basis is projected, selecting only designated parts of the input field. The selected field is then projected by a lens onto a detector where the intensity $y_i$ is measured by integrating the intensity over the whole detector surface.
    \textbf{B)} Different measurement bases of size $N = 16$ and the projection of the second measurement vector onto a $(4 \times 4)$ 2d grid for sampling with a SLM or DMD.  } 
    \label{fig:setup}
\end{figure}

Conventional cameras measure the spatial intensity distribution of light using a two-dimensional array of detectors, such as CCD or CMOS sensors. In contrast, Single Pixel Imaging (SPI) employs a single photodetector (bucket detector) that integrates the intensity of a spatially modulated light field. A schematic of a typical experimental setup is shown in Fig.~\ref{fig:setup}A. A collimated input field $x$ is incident on a DMD or SLM, where it is sequentially modulated by an orthogonal measurement basis $\Phi$. The modulated field is projected onto a detector using a lens with focal length $f$, which can be arranged either in an imaging configuration or with the detector placed in the focal plane, provided that the entire field is collected. The distances $s$ (SLM to lens) and $i$ (lens to detector) satisfy the imaging equation
$1/f = 1/i + 1/s$.
A CCD camera may also be used as a bucket detector by integrating the signal over all pixels.

The measurement basis consists of orthogonal row vectors $\Phi_i$ that are mapped onto a two-dimensional $m \times n$ superpixel grid of the SLM. Typically, a square subsection of size $n \times n$ is used, where $n = \alpha \sqrt{N}$ with $\alpha \in \mathbb{N}$ to ensure a direct mapping of the basis elements (Fig.~\ref{fig:setup}B). Each basis vector $\Phi_i$ is reshaped row-wise into a two-dimensional pattern and projected onto the SLM. The corresponding detector signal is given by $ y_i = \Phi_i \cdot x.$
By repeating this procedure for all basis vectors, the input field can be reconstructed as
\begin{equation}
    x = \Phi^{-1} y.
    \label{eq:SPI}
\end{equation}

The measurement basis does not need to be implemented on a two-dimensional grid, but it can also be applied to one-dimensional structures, such as a ring, for example when sampling Bessel beam–like light fields \cite{Scheidt:Bessel}. In interferometric SPI setups, such as those used to measure complex wavefronts, interference effects arising from the size of the measurement modes can limit reconstruction performance. This is because the pixels on a DMD or SLM are typically at least an order of magnitude larger than the sampled image source, which can introduce phase aberrations that appear in amplitude reconstructions \cite{Scheidt:24}.

\subsection{Measurement Bases}

A measurement basis $\Phi$ is defined as an orthogonal matrix whose vectors satisfy $\Phi_i \cdot \Phi_j = \delta_{ij}$. Commonly used bases in SPI include the canonical (identity), random Gaussian, and Hadamard bases \cite{reviewsinglepixel}. The canonical basis corresponds to the identity matrix and directly samples individual spatial elements.

The Hadamard basis consists of entries $\pm1$ and is defined for sizes $N = 2^n$. It is constructed recursively using the Kronecker product of the $2 \times 2$ Hadamard matrix $H_2 = \begin{pmatrix}1 & 1 \\ 1 & -1\end{pmatrix}$, yielding the natural-ordered Hadamard matrix (Fig.~\ref{fig:setup}B). The Hadamard matrix satisfies $H H^{-1} = N I$. Alternative orderings include the Walsh-ordered Hadamard basis, obtained by sorting the Walsh functions according to increasing frequency \cite{Walsh}, and the cake-cutting ordering, which is based on partitioning the reshaped basis vectors into regions of increasing spatial complexity \cite{CakeCutting}.

Randomized binary Gaussian bases, with elements $\{0,1\}$ or $\{\pm1\}$, are also frequently employed. Unlike the Hadamard basis, these matrices allow arbitrary dimensions and are less sensitive to ordering effects in compressive sensing, as each measurement samples a broad range of spatial frequencies.

Implementing Hadamard patterns on an SLM or DMD requires addressing both positive and negative values. A straightforward approach separates the basis into positive ($H^+$) and negative ($H^-$) components, requiring twice the number of measurements. This redundancy can be avoided by exploiting the structure of the natural Hadamard matrix, where the first row contains only $+1$ entries. Using the relation \cite{SPI_HadFormula}
\begin{equation}
    H_{2\ldots N} = 2 H^+_{2\ldots N} - H_1^+,
    \label{eq:Had_exp}
\end{equation}
all measurements can be reconstructed from a single binary acquisition, significantly reducing measurement time. An analogous strategy is used in interferometric wavefront sensing by encoding $\pm1$ values as phase shifts of $0$ and $\pi$, respectively, such that subtraction is performed optically by the Fourier-transforming lens \cite{T_Mat3,ota,Scheidt:23}.

Notice, that depending on the device, basis implementation might differ. While the basis patterns can be directly projected onto a DMD or amplitude modulating SLM, phase-only SLMs require the superpositon of a phase mask that enables amplitude modulation \cite{Scheidt:23,Scheidt:24}.

Memory requirements pose an additional challenge for large Hadamard matrices. Storing a full $N \times N$ matrix with $\pm1$ entries requires substantial memory (e.g., 16.7~GB for $N=4096$ using 8-bit integers). Binary representations combined with Eq.~\ref{eq:Had_exp} reduce memory consumption by a factor of eight, though large-scale implementations still demand significant computational resources.

\subsection{Compressive Sensing}

\subsubsection{Basis Pursuit Reconstruction}

Since the full measurement basis contains redundant information, Eq.~\ref{eq:SPI} can be solved using CS with a reduced number of measurements $M < N$. Successful reconstruction requires incoherence of the measurement matrix and sufficient sparsity $K$ of the signal representation, with
\begin{equation}
    M \approx K \log(N/K).
    \label{eq:sparsity}
\end{equation}
Hadamard and random Gaussian bases satisfy the incoherence requirement in optical systems \cite{Tajahuerce:14}. The reconstruction problem is formulated using a sparsifying transform $\Psi$, such as the discrete cosine transform (DCT), yielding $\Theta = \Phi \Psi$. The signal is recovered by solving the $\ell_1$-minimization problem
\begin{equation}
    \min \|s\|_1 \quad \text{subject to} \quad y = \Theta s,
    \label{eq:CS_rec}
\end{equation}
which is implemented using the SPGL1 basis pursuit denoising algorithm \cite{BergFriedlander:2008,spgl1site}. The reconstructed signal is transformed back to the original domain using the inverse DCT.

\subsubsection{Deep Learning Reconstruction}

For deep learning–based reconstruction, the CS problem is reformulated as a supervised regression task \cite{YAO2019483}:
\begin{equation}
    \mathbf{W}^f = \arg\min_{\mathbf{W}} \|\mathbf{x} - \mathbf{W}\mathbf{y}\|_2^2.
    \label{eq:CS_Linear}
\end{equation}
A linear mapping network $\mathcal{G}^f$ consisting of a single fully connected layer is trained to infer the optimal mapping matrix $\mathbf{W}^f$ using paired training samples $(\mathbf{y}_i,\mathbf{x}_i)$. The loss function is defined as
\begin{equation}
    L = \frac{1}{N} \sum_{i=1}^N \|\mathbf{x}_i - \mathcal{G}^f(\mathbf{y}_i)\|_2^2.
\end{equation}
The network output is reshaped into a $32 \times 32$ reconstruction \cite{YAO2019483}.

Training and evaluation are performed using the CIFAR-10 dataset, which contains 60,000 images of size $32 \times 32$ \cite{cifar10}. CS measurements are generated using natural Hadamard, Walsh-ordered Hadamard, cake-cutting ordered Hadamard, and random Gaussian sensing matrices across compression ratios ($cr$) ranging from 1\% to 50\%. The networks are implemented in TensorFlow and trained for $50$ epochs using the Adam optimizer \cite{KingBa15}, with a learning rate of $0.1$ that is adaptively reduced by a factor of $0.95$ upon convergence after 10 epochs.
The performance of the deep learning approach can be further improved by increasing the number of training epochs, enlarging the training dataset, employing batch optimization, and using more complex neural network architectures. Note, that this process shall be applied to each trained model, depending on the compression ratio and sampling basis. However, because the code is provided as a publicly accessible \href{https://colab.research.google.com/drive/1d0hrh3v9VdLLxFooUzy88cT48RY8_TWf?usp=sharing}{Google Colab notebook}, the training parameters are intentionally constrained to allow execution without GPU acceleration or additional computational resources.

\section{Comparison of Compressive Sensing Algorithms}\label{sec:CS}

This section examines the influence of different measurement bases on the reconstruction performance of deterministic compressive sensing algorithms, using both a simple implementation of the basis pursuit that uses iterative soft thresholding algorithm (ISTA) \cite{Liu2015ASA} and the SPGL1 package \cite{spgl1site}. For this purpose, an image of the cifar-10 dataset \cite{cifar10} is sampled at different compression ratios $cr = M/N $ $ [1,2,5,10,20,25,30,40,50,60,75,100]\%$, and the results are compared to the original image for various orderings of the Hadamard basis as well as the random Gaussian basis.
Typical metrics for evaluating reconstruction quality are the Root Mean Square Error (RMSE), Peak Signal-to-Noise Ratio (PSNR), and Structural Similarity Index (SSIM) between a reconstructed image $\hat{x}$ and a reference image $x$ \cite{databook}:

\[
\begin{aligned}
\mathrm{RMSE} &= \sqrt{\frac{1}{N} \sum_{i=1}^{N} \left( x_i - \hat{x}_i \right)^2 }, \\
\mathrm{PSNR} &= 10 \cdot \log_{10} \frac{\max(x)^2}{\mathrm{MSE}}, \quad 
\mathrm{MSE} = \frac{1}{N} \sum_{i=1}^{N} \left( x_i - \hat{x}_i \right)^2, \\
\mathrm{SSIM}(x, \hat{x}) &= \frac{(2 \mu_x \mu_{\hat{x}} + C_1)(2 \sigma_{x\hat{x}} + C_2)}{(\mu_x^2 + \mu_{\hat{x}}^2 + C_1)(\sigma_x^2 + \sigma_{\hat{x}}^2 + C_2)},
\end{aligned}
\]

where $\mu$ denotes the mean intensity, $\sigma^2$ the variance, $\sigma_{x\hat{x}}$ the covariance, and $C_1$, $C_2$ are small constants for stability.

\subsection{Full-Basis Sampling and Reconstruction}

\begin{figure}[h]
    \centering
    \includegraphics[width=0.75\linewidth]{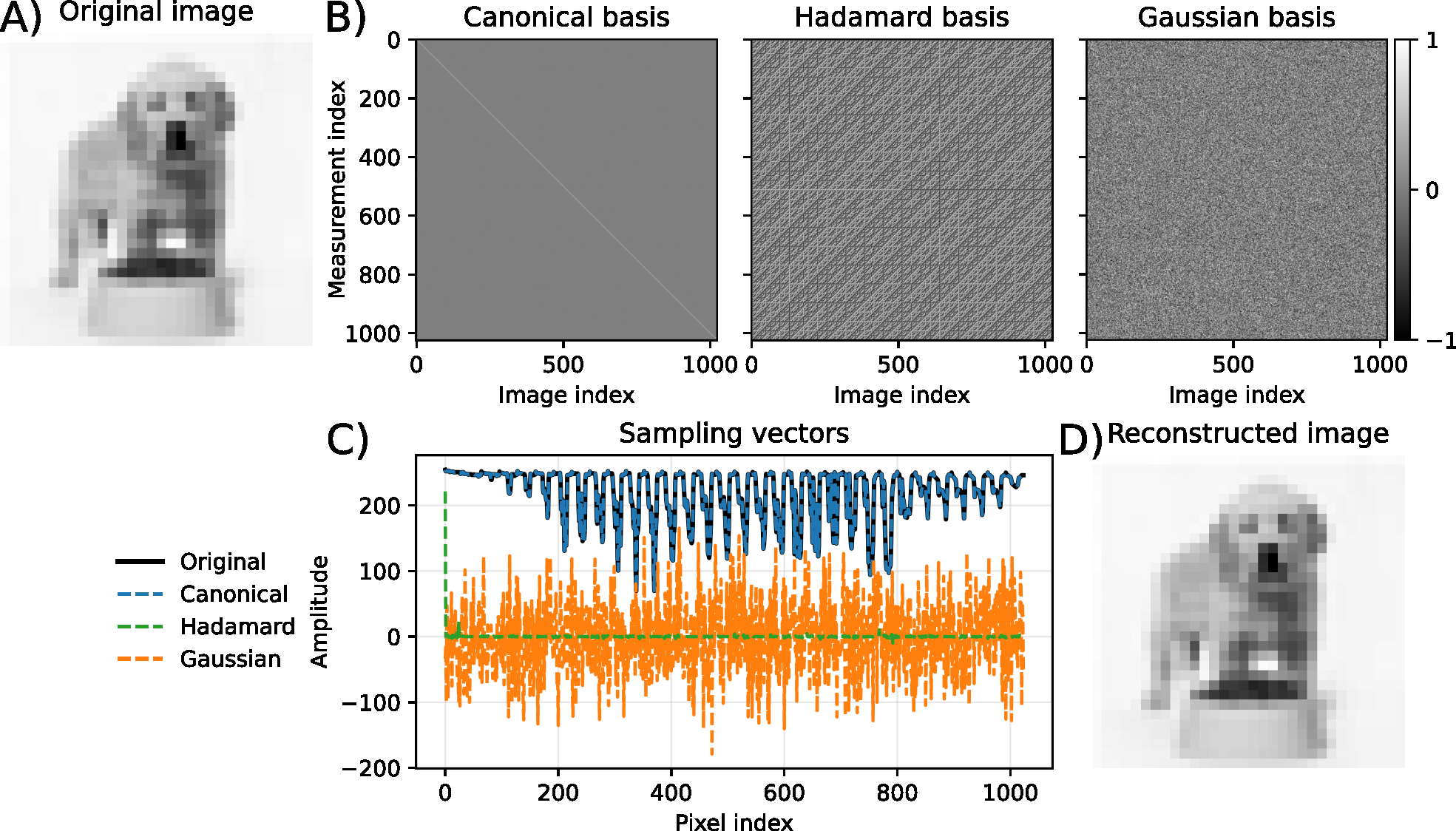}
    \caption{\textbf{A)} Original image. \textbf{B)} Measurement bases: Canonical, Hadamard, and random Gaussian. \textbf{C)} One-dimensional vector representations of the original image (black) and sampled signals using the canonical (blue), Hadamard (green), and random Gaussian (orange) bases. \textbf{D)} Reconstructed image using Eq.~\ref{eq:SPI}.}
    \label{fig:reconstruction}
\end{figure}

Figure~\ref{fig:reconstruction}A shows a $32 \times 32$ dog image $x$, which is digitally sampled using the canonical, Hadamard, and random Gaussian bases, illustrated in inset B. The sampling process is mathematically equivalent to multiplying the one-dimensional image vector $x$ of size $N \times 1$ by the measurement matrix $\Phi$:
\begin{equation}
    y = \Phi \cdot x.
\end{equation}

The resulting 1D measurement vectors $y$ are shown in Fig.~\ref{fig:reconstruction}C. The original image is depicted as a solid black line, while the canonical, Hadamard, and Gaussian sampled vectors are shown as blue, green, and orange dashed lines, respectively. The canonical basis reproduces the original image exactly, as it independently samples each spatial element. The reconstructed image obtained from these measurements, using Eq.~\ref{eq:SPI}, is displayed in Fig.~\ref{fig:reconstruction}D. In practice, the inverse of $\Phi$ can be computed using a pseudo-inverse function, such as \texttt{pinv}, in most programming languages.
Note that when using the full measurement basis for reconstruction, the original image can be recovered. However, in experimental setups, factors such as illumination and acquisition noise, as well as optical aberrations, may reduce the quality of the reconstructed image.

\subsection{Influence of the basis ordering}

\begin{figure}[h]
    \centering
    \includegraphics[width=\linewidth]{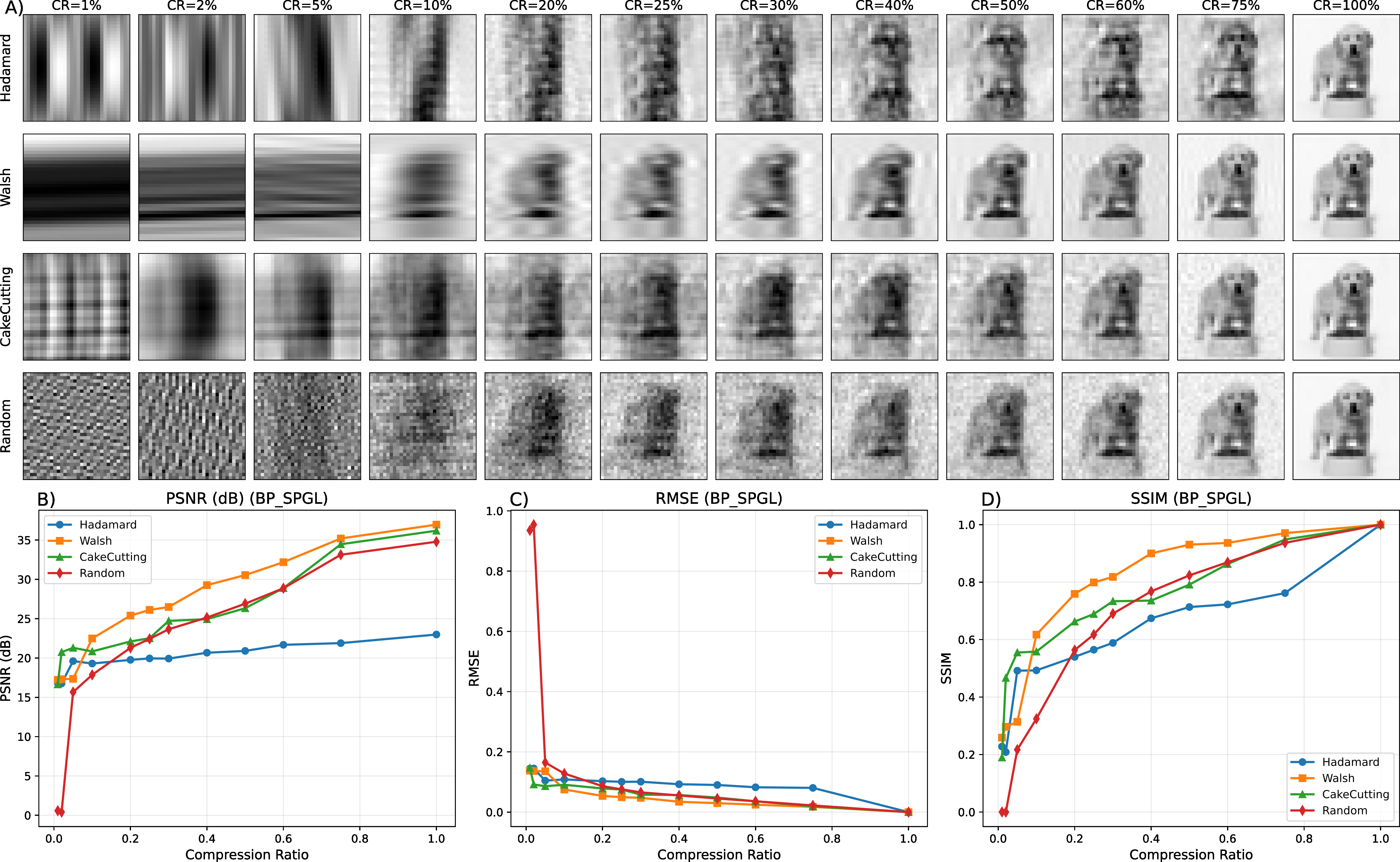}
    \caption{\textbf{A)} Reconstructed images using the SPGL1 basis pursuit algorithm, with increasing compression ratio along the columns. Rows show, from top to bottom, the natural Hadamard, Hadamard-Walsh, Cake-Cutting, and random Gaussian orderings. \textbf{B)} PSNR, \textbf{C)} RMSE, and \textbf{D)} SSIM of the reconstructed images for the natural Hadamard (blue), Hadamard-Walsh (orange), Cake-Cutting (green), and random Gaussian (red) bases as a function of the compression ratio.}
    \label{fig:ordering}
\end{figure}

Figure~\ref{fig:ordering}A shows the reconstructed dog image using the basis pursuit algorithm from SPGL1 for the natural Hadamard, Hadamard-Walsh, Cake-Cutting ordered Hadamard, and random Gaussian bases (rows) across different compression ratios (columns). The general shape of the dog is already visible at compression ratios of $20\%$, independent of the measurement basis. Increasing the compression ratio reveals finer details and features.  

Reconstructions using the natural Hadamard basis exhibit vertical repetitions due to missing frequency components in the sampling process, noticeable up to a compression ratio of $75\%$. The Hadamard-Walsh basis provides a good reconstruction of the shape at compression ratios of $20$–$25\%$, as it effectively acts as a low-pass filter in the sampling domain according to $N_{\mathrm{res}} = 2^{\text{ceil}( \log_2(M))}$ \cite{Scheidt:Bessel,HadOrder}. Increasing the number of measurements $M$ raises the maximum resolvable sampling frequency in the reconstructed images.  

The Cake-Cutting and random Gaussian bases sample multiple frequency domains simultaneously, resulting in noisier reconstructions. This noise decreases as $M$ increases.

Figures~\ref{fig:ordering}B--D plot PSNR, RMSE, and SSIM as a function of the compression ratio for the natural Hadamard (blue), Hadamard-Walsh (orange), Cake-Cutting (green), and random Gaussian (red) bases. In general, all metrics show a strong increase in the range $cr = 10$–$20\%$, after which the curves approach a linear trend. Across all metrics, the natural Hadamard basis performs worst, while the Hadamard-Walsh basis performs best. The Cake-Cutting and random Gaussian bases yield similar performance.

\subsection{Influence of Reconstruction Algorithm}

Since the Hadamard-Walsh ordering yields the best performance among the measurement bases for deterministic reconstruction, different compressive sensing reconstruction algorithms are compared: a basic implementation of the basis pursuit algorithm, the basis pursuit algorithm provided by SPGL1, and the least absolute shrinkage and selection operator (LASSO) algorithm from the SPGL1 package. The latter corresponds to a basis pursuit denoising approach that incorporates an $\ell_1$ regularization term, whose strength is controlled by the parameter $\lambda$.

\begin{figure}[h]
    \centering
    \includegraphics[width=\linewidth]{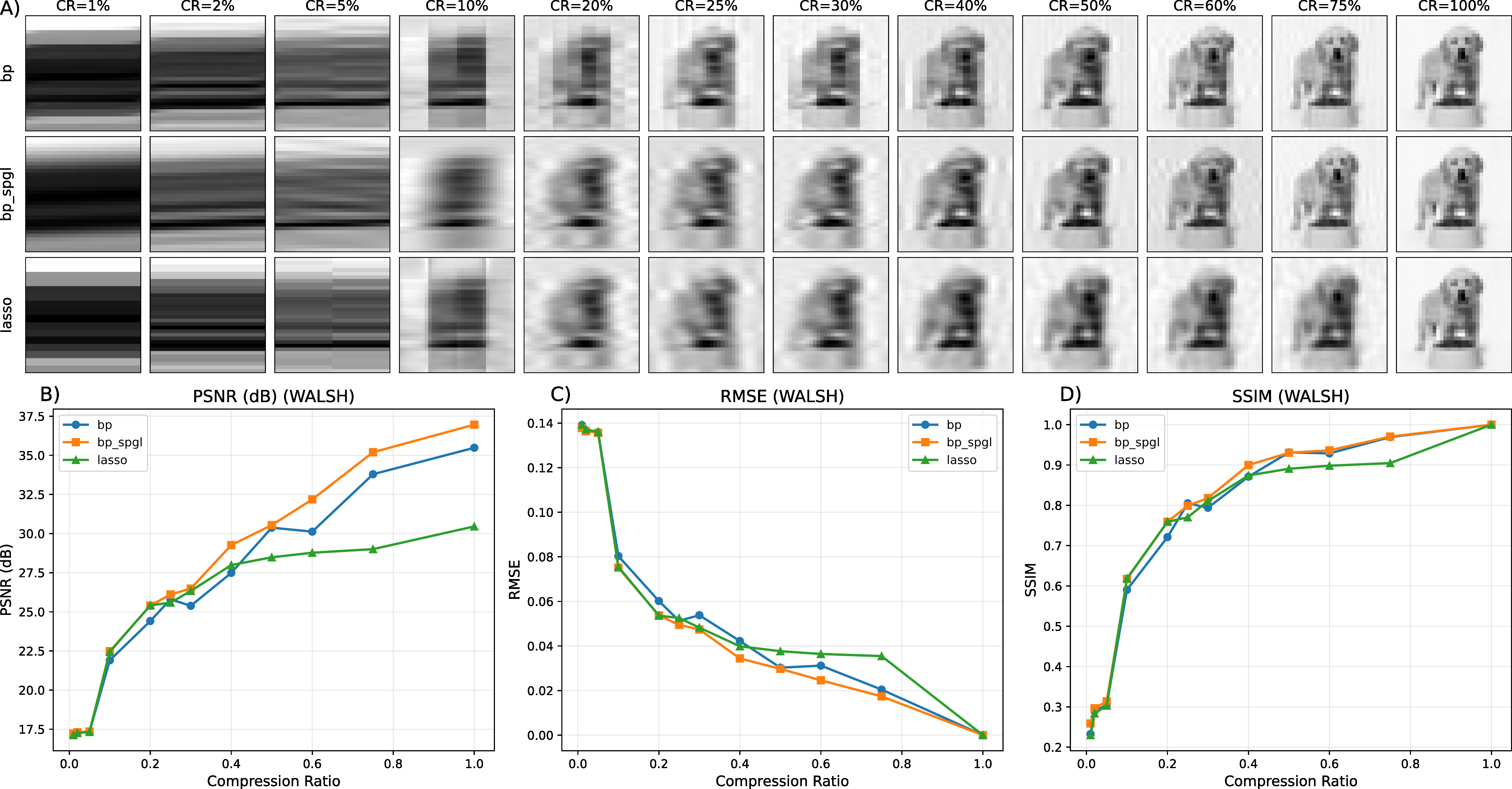}
    \caption{\textbf{A)} Reconstructed images sampled with the Hadamard-Walsh basis with increasing compression ratio along the columns. Rows show, from top to bottom, the basic basis pursuit, SPGL1 basis pursuit, and SPGL1 LASSO algorithms. \textbf{B)} PSNR, \textbf{C)} RMSE, and \textbf{D)} SSIM of the reconstructed images for the basic basis pursuit (blue), SPGL1 basis pursuit (orange), and SPGL1 LASSO (green) algorithms as a function of the compression ratio.}
    \label{fig:algorithm}
\end{figure}

Figure~\ref{fig:algorithm}A shows the reconstructed images with increasing compression ratio along the columns. The rows display, from top to bottom, the basic basis pursuit implementation, the SPGL1 basis pursuit algorithm, and the SPGL1 LASSO algorithm. Similar to the results shown in Fig.~\ref{fig:ordering}, the overall shape of the dog becomes visible at compression ratios of $cr = 20$--$25\%$. Further increasing the compression ratio reveals higher spatial-frequency details of the image.  

The basic basis pursuit implementation exhibits rectangular superpixel structures up to $cr = 50\%$, which are related to the limited number of resolvable basis elements $N_{\mathrm{res}}$. In contrast, the SPGL1-based algorithms include built-in regularization that smooths transitions between pixel boundaries, reducing the visibility of these superpixels. Nevertheless, superpixel regimes remain apparent at low compression ratios of $cr = 1$--$10\%$. The SPGL1 LASSO algorithm produces slightly blurred reconstructions, which effectively preserve low-frequency image features, particularly in the background, resulting in a visually cleaner overall image.

However, the smoothing introduced by regularization results in a reduced similarity performance, as shown in Figs.~\ref{fig:algorithm}B--D (basic implementation: blue; SPGL1 basis pursuit: orange; SPGL1 LASSO: green). For example, the PSNR is approximately 5~dB lower at $cr = 50\%$ and decreases by up to 7.5~dB at $cr = 75\%$. Similarly, the RMSE increases by about 2 percentage points, while the SSIM decreases by approximately 5 percentage points.  

Despite the absence of explicit regularization in the basic basis pursuit implementation, both basis pursuit approaches exhibit very similar performance, with only minor differences. Notably, the impact of regularization becomes apparent at $cr \approx 40\%$ for PSNR and SSIM, and at $cr \approx 30\%$ for RMSE. Overall, the SPGL1 basis pursuit algorithm provides the highest reconstruction fidelity among the evaluated methods.

To facilitate hands-on practice, the accompanying \href{https://colab.research.google.com/drive/1BOQHEbJbRKe9fbTBnLtMMlTHFeOjXtF8?usp=sharing}{notebook}  provides code to read and process experimentally acquired datasets from \cite{Scheidt_2025}, allowing evaluation in a manner analogous to the synthetic CIFAR-10 images. Figures are automatically generated during this analysis, enabling readers to visualize the results immediately. Additionally, the notebook supports the use of other images from the CIFAR-10 dataset, allowing exploration of the algorithms under different conditions.

\section{Deep Learning Reconstruction using a Linear Network}\label{sec:DL}

\begin{figure}[h]
    \centering
    \includegraphics[width=0.75\linewidth]{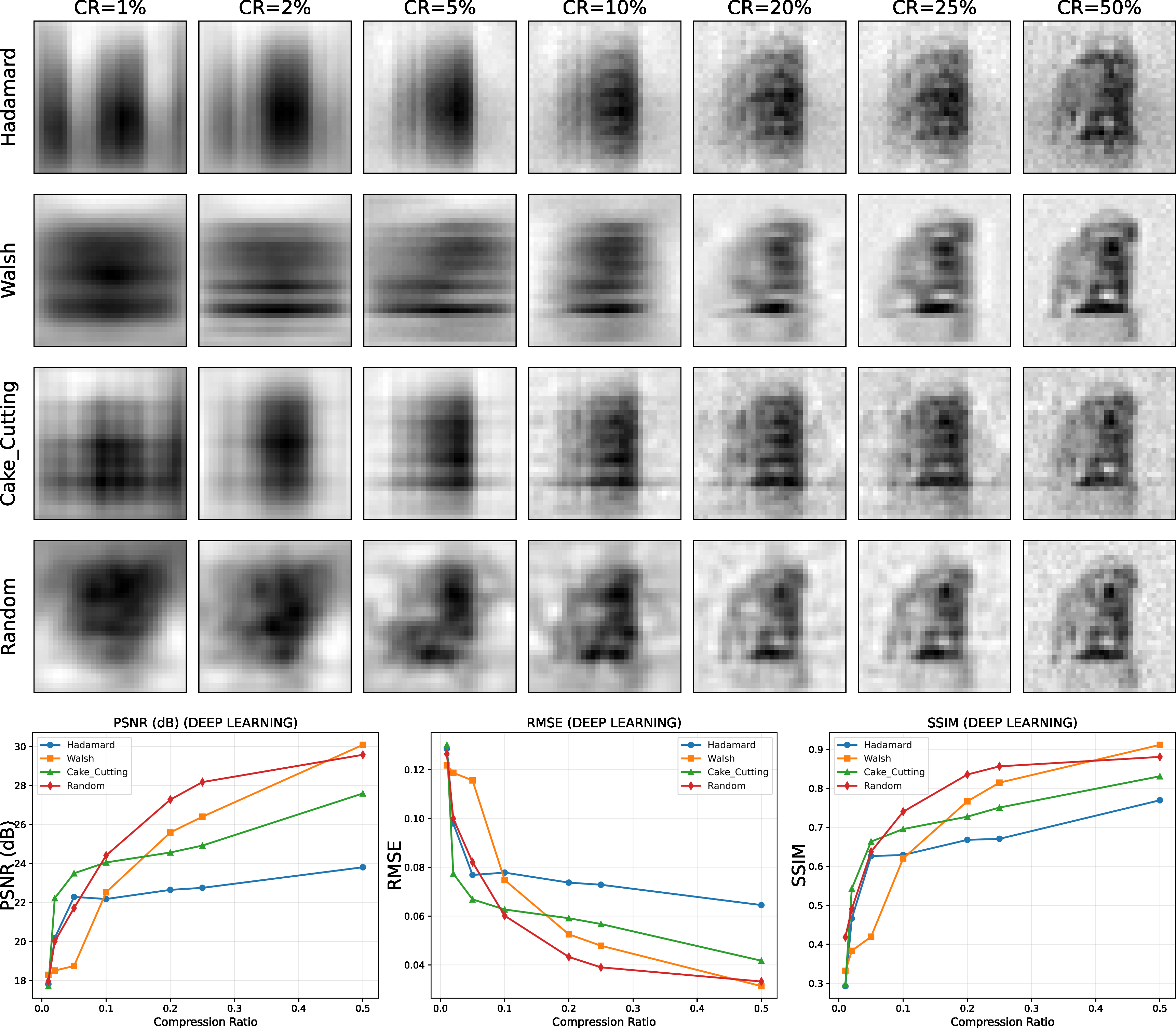}
    \caption{\textbf{A)} Reconstructed images using trained linear neural networks, with increasing compression ratio along the columns. Rows show, from top to bottom, the natural Hadamard, Hadamard-Walsh, Cake-Cutting, and random Gaussian orderings. \textbf{B)} PSNR, \textbf{C)} RMSE, and \textbf{D)} SSIM of the reconstructed images for the natural Hadamard (blue), Hadamard-Walsh (orange), Cake-Cutting (green), and random Gaussian (red) bases as a function of the compression ratio.}
    \label{fig:LinearModel}
\end{figure}

The linear networks were trained using a learning rate of $0.01$, batch size of $n_{batch} = 256$, Adam optimizers $\beta_1 = 0.9$, $\beta_2 = 0.99$ and $\epsilon = 0.1$ for $n_{epochs} = 50$ epochs.
Deep learning–based reconstruction using a simple linear neural network is shown in Fig.~\ref{fig:LinearModel}A for compression ratios $cr = [1, 2, 5, 10, 20, 25, 50]\%$ along the columns. The rows display, from top to bottom, the natural Hadamard, Hadamard-Walsh, Cake-Cutting, and random Gaussian basis orderings. In contrast to deterministic compressive sensing algorithms, each reconstruction requires a separately trained model for the chosen compression ratio and measurement basis.

Compared to the deterministic reconstructions in Fig.~\ref{fig:ordering}A, the main features of the dog image are already visible at a compression ratio of $cr = 5\%$ for the random Gaussian basis and at $cr = 10\%$ for the Cake-Cutting ordering. At $cr = 20\%$, finer image details become apparent. The Cake-Cutting ordering, however, exhibits artifacts related to its frequency sampling characteristics, and the background remains comparatively noisy. These bases previously performed worse than the Hadamard-Walsh ordering, which is also reflected here. The frequency-doubling artifacts observed for the natural Hadamard ordering are still present but are significantly reduced at $cr = 50\%$ compared to the deterministic reconstruction results. Similarly, the low-pass–filter-related blurring associated with the Hadamard-Walsh basis is reduced.

Figures~\ref{fig:LinearModel}B--D show PSNR, RMSE, and SSIM as a function of compression ratio for the natural Hadamard (blue), Hadamard-Walsh (orange), Cake-Cutting (green), and random Gaussian (red) bases. All metrics exhibit a steep increase at low compression ratios. In particular, the random Gaussian basis outperforms the other bases by more than 5\% from $cr = 10\%$ up to $cr = 50\%$, where the Hadamard-Walsh basis slightly surpasses it. This behavior arises because the random Gaussian basis contains a broad range of spatial frequency components in the compressed measurements, whereas the Hadamard basis selectively samples specific frequencies. The increased information content enables the fully connected neural network to learn more effective weighting, thereby improving reconstruction performance.

All networks are trained under identical conditions. Reconstruction performance can be significantly improved by optimizing training parameters such as learning rate, learning rate scheduling, batch size, and activation functions. Typically, this involves an initial parameter sweep over a limited number of epochs, followed by extended training using the optimal parameter set and a larger training dataset. Such optimization, however, requires substantially greater computational resources and is therefore best performed using GPU acceleration.

For the reconstruction of experimental data, the inclusion of artificial noise during training is strongly recommended. Training solely on noise-free synthetic data can lead to artifacts when applied to experimental measurements. In many cases, directly training the network on experimental data yields superior performance by adapting the model to the specific noise characteristics of the setup. This approach, however, is considerably more time-consuming and demands the acquisition of large amount of experiment specific data.

Readers are strongly encouraged to copy the code provided in the \href{https://colab.research.google.com/drive/1d0hrh3v9VdLLxFooUzy88cT48RY8_TWf?usp=sharing}{training notebook} and perform parameter optimization. An example is included to demonstrate the effect of such an optimized neural network. Optimizing the learning rate to $lr  = 0.005$, the  batch size to $n_{batch} = 64$ and the Adam parameters $\beta_1 = 0.99$, $\beta_2'= 0.99$ and $\epsilon = 0.1$ and training for $n_{epochs} = 200$ epochs yields an improvement of $0.5\%$.

\section{Conclusion}

This tutorial provides a comprehensive introduction to single pixel imaging, demonstrating the experimental implementation of sampling bases using devices such as spatial light modulators (SLMs) and digital micromirror devices (DMDs). The concept of compressive sensing is introduced, and the influence of different sampling bases and deterministic reconstruction algorithms is systematically discussed.

For experimentally acquired data, the Hadamard–Walsh basis in combination with a basis pursuit denoising algorithm, such as those provided by the SPGL1 package, is shown to be well suited. Acting as a low-pass filter, the Hadamard–Walsh basis preserves the global structure of the image at compression ratios of approximately $20\%$. This property is particularly advantageous for optical experiments with low perturbations, where wavefronts and optical fields vary smoothly.

Finally, compressive sensing reconstruction using a simple linear neural network is demonstrated for single pixel imaging data across different measurement bases and compression ratios. In contrast to deterministic reconstruction algorithms, randomized Gaussian measurement matrices perform best in this context, as they contain all spatial frequency components in the compressed signal and yield good reconstruction quality already at compression ratios as low as $10\%$. However, neural networks are inherently task-specific and require dedicated training for each measurement configuration, limiting their general adaptability.

To support reproducibility and hands-on exploration, Jupyter notebooks  implementing all algorithms used to generate the results presented in this tutorial are provided (\href{https://colab.research.google.com/drive/1BOQHEbJbRKe9fbTBnLtMMlTHFeOjXtF8?usp=sharing}{algorithms and results} and \href{https://colab.research.google.com/drive/1d0hrh3v9VdLLxFooUzy88cT48RY8_TWf?usp=sharing}{neuronal network training}). Readers are encouraged to use these resources to further explore and adapt single pixel imaging techniques for their own applications.

\bibliographystyle{elsarticle-harv} 
\bibliography{bibliography}

@article{Zupancic:16,
author = {Philip Zupancic and Philipp M. Preiss and Ruichao Ma and Alexander Lukin and M. Eric Tai and Matthew Rispoli and Rajibul Islam and Markus Greiner},
journal = {Opt. Express},
number = {13},
pages = {13881--13893},
publisher = {OSA},
title = {Ultra-precise holographic beam shaping for microscopic quantum control},
volume = {24},
month = {06},
year = {2016},
doi = {10.1364/OE.24.013881},
}

@article{Tajahuerce:14,
author = {Enrique Tajahuerce and Vicente Dur\'{a}n and Pere Clemente and Esther Irles and Fernando Soldevila and Pedro Andr\'{e}s and Jes\'{u}s Lancis},
journal = {Opt. Express},
number = {14},
pages = {16945--16955},
publisher = {OSA},
title = {Image transmission through dynamic scattering media by single-pixel photodetection},
volume = {22},
month = {07},
year = {2014},
doi = {10.1364/OE.22.016945},
}

@article{Scheidt:23,
author = {Dennis Scheidt and Pedro A. Quinto-Su},
journal = {J. Opt. Soc. Am. A},
number = {1},
pages = {45--52},
publisher = {Optica Publishing Group},
title = {Comparison between Hadamard and canonical bases for in situ wavefront correction and the effect of ordering in compressive sensing},
volume = {40},
month = {01},
year = {2023},
doi = {10.1364/JOSAA.473940},
}

@article{Scheidt:24,
author = {Dennis Scheidt and Pedro A. Quinto-Su},
journal = {Opt. Lett.},
number = {9},
pages = {2381--2384},
publisher = {Optica Publishing Group},
title = {Spatial resolution limit of single pixel imaging of complex light fields},
volume = {49},
month = {May},
year = {2024},
doi = {10.1364/OL.519587},
}

@article{SinglePixel,
  author  = {Liu, Ruifeng and Zhao, Shupeng and Zhang, Pei and Gao, Hong and Li, Fuli},
  title   = {Complex wavefront reconstruction with single-pixel detector},
  journal = {Applied Physics Letters},
  year    = {2019},
  volume  = {114},
  number  = {16},
  pages   = {161901},
}

@article{SPI_HadFormula,
author = {Daixuan, Wu and Jiawei, Luo and Guoqiang, Huang and Feng, Yuanhua and Xiaohua, Feng and Shen, Yuecheng and Li, Zhao-Hui},
year = {2020},
month = {12},
pages = {},
title = {Imaging biological tissue with high-throughput single-pixel compressive holography},
doi = {10.21203/rs.3.rs-129598/v1},
journal = {Nat. Commun.}
}

@ARTICLE{Duarte,
  author={Duarte, Marco F. and Davenport, Mark A. and Takhar, Dharmpal and Laska, Jason N. and Sun, Ting and Kelly, Kevin F. and Baraniuk, Richard G.},
  journal={IEEE Signal Processing Magazine}, 
  title={Single-pixel imaging via compressive sampling}, 
  year={2008},
  volume={25},
  number={2},
  pages={83-91},
  doi={10.1109/MSP.2007.914730}}

@article{PhysRevLett.74.3600,
  title = {Observation of Two-Photon ``Ghost'' Interference and Diffraction},
  author = {Strekalov, D. V. and Sergienko, A. V. and Klyshko, D. N. and Shih, Y. H.},
  journal = {Phys. Rev. Lett.},
  volume = {74},
  issue = {18},
  pages = {3600--3603},
  numpages = {0},
  year = {1995},
  month = {May},
  publisher = {American Physical Society},
  doi = {10.1103/PhysRevLett.74.3600}
}

@article{PhysRevA.52.R3429,
  title = {Optical imaging by means of two-photon quantum entanglement},
  author = {Pittman, T. B. and Shih, Y. H. and Strekalov, D. V. and Sergienko, A. V.},
  journal = {Phys. Rev. A},
  volume = {52},
  issue = {5},
  pages = {R3429--R3432},
  numpages = {0},
  year = {1995},
  month = {Nov},
  publisher = {American Physical Society},
  doi = {10.1103/PhysRevA.52.R3429},
  }

@article{PhysRevA.86.041803,
  title = {Single-pixel digital ghost holography},
  author = {Clemente, Pere and Dur\'an, Vicente and Tajahuerce, Enrique and Torres-Company, Victor and Lancis, Jesus},
  journal = {Phys. Rev. A},
  volume = {86},
  issue = {4},
  pages = {041803},
  numpages = {4},
  year = {2012},
  month = {Oct},
  publisher = {American Physical Society},
  doi = {10.1103/PhysRevA.86.041803},
  }

@article{Padgett_CGI,
author = {Padgett, Miles and Boyd, Robert},
year = {2017},
month = {06},
pages = {20160233},
title = {An introduction to ghost imaging: Quantum and classical},
volume = {375},
journal = {Philosophical Transactions of The Royal Society A Mathematical Physical and Engineering Sciences},
doi = {10.1098/rsta.2016.0233}
}

@article{GONG2022108140,
title = {Performance comparison of computational ghost imaging versus single-pixel camera in light disturbance environment},
journal = {Optics \& Laser Technology},
volume = {152},
pages = {108140},
year = {2022},
issn = {0030-3992},
doi = {https://doi.org/10.1016/j.optlastec.2022.108140},
author = {Wenlin Gong}
}

@article{Sephton:23_GhostImagingPhase,
author = {Bereneice Sephton and Isaac Nape and Chan\'{e} Moodley and Jason Francis and Andrew Forbes},
journal = {Optica},
number = {2},
pages = {286--291},
publisher = {Optica Publishing Group},
title = {Revealing the embedded phase in single-pixel quantum ghost imaging},
volume = {10},
month = {02},
year = {2023},

doi = {10.1364/OPTICA.472980},
}

@article{Clemente:13,
author = {Pere Clemente and Vicente Dur\'{a}n and Enrique Tajahuerce and Pedro Andr\'{e}s and Vicent Climent and Jes\'{u}s Lancis},
journal = {Opt. Lett.},
number = {14},
pages = {2524--2527},
publisher = {OSA},
title = {Compressive holography with a single-pixel detector},
volume = {38},
month = {07},
year = {2013},
url = {http://www.osapublishing.org/ol/abstract.cfm?URI=ol-38-14-2524},
doi = {10.1364/OL.38.002524},
}

@article{Horisaki:17_2,
author = {Ryoichi Horisaki and Hiroaki Matsui and Jun Tanida},
journal = {Appl. Opt.},
number = {14},
pages = {4085--4089},
publisher = {OSA},
title = {Single-pixel compressive diffractive imaging with structured illumination},
volume = {56},
month = {05},
year = {2017},
url = {http://www.osapublishing.org/ao/abstract.cfm?URI=ao-56-14-4085},
doi = {10.1364/AO.56.004085},
}

@article{Howland:14,
author = {Gregory A. Howland and Daniel J. Lum and John C. Howell},
journal = {Opt. Express},
number = {16},
pages = {18870--18880},
publisher = {OSA},
title = {Compressive wavefront sensing with weak values},
volume = {22},
month = {08},
year = {2014},
url = {http://www.osapublishing.org/oe/abstract.cfm?URI=oe-22-16-18870},
doi = {10.1364/OE.22.018870},
}

@Article{T_Mat3,
  author  = {A Liutkus; D Martina; S Popoff; et al},
  journal = {Scientific reports},
  title   = {Imaging with nature: Compressive imaging using a multiply scattering medium},
  year    = {2014},
  month   = {01},
  number  = {5552},
  pages   = {5552},
  volume  = {4}
}

@article{Dutta:19,
author = {Rahul Dutta and Silvestre Manzanera and Adri\'{a}n Gamb\'{i}n-Regadera and Esther Irles and Enrique Tajahuerce and Jes\'{u}s Lancis and Pablo Artal},
journal = {Biomed. Opt. Express},
number = {8},
pages = {4159--4167},
publisher = {OSA},
title = {Single-pixel imaging of the retina through scattering media},
volume = {10},
month = {08},
year = {2019},
url = {http://www.osapublishing.org/boe/abstract.cfm?URI=boe-10-8-4159},
doi = {10.1364/BOE.10.004159},
}

@article{Wang:Mouse,
author = {Wang, Daifa and Zhou, Haojiang and Brake, Joshua and Ruan, Haowen and Jang, Mooseok and Yang, And},
year = {2015},
month = {08},
pages = {728-735},
title = {Focusing through dynamic tissue with millisecond digital optical phase conjugation},
volume = {2},
journal = {Optica},
doi = {10.1364/OPTICA.2.000728}
}

@article{Boniface:19,
author = {Antoine Boniface and Baptiste Blochet and Jonathan Dong and Sylvain Gigan},
journal = {Optica},
number = {11},
pages = {1381--1385},
publisher = {OSA},
title = {Noninvasive light focusing in scattering media using speckle variance optimization},
volume = {6},
month = {11},
year = {2019},
url = {http://www.osapublishing.org/optica/abstract.cfm?URI=optica-6-11-1381},
doi = {10.1364/OPTICA.6.001381},
}

@Article{HadOrder,
  author  = {Vaz P G ;Amaral D ; Requicha Ferreira L F; et al},
  journal = {Optics Express},
  title   = {Image quality of compressive single-pixel imaging using different Hadamard orderings},
  year    = {2020},
  month   = {04},
  number  = {888},
  volume  = {28}
}

@Article{HadOrigami,
  author         = {Yu, Wen-Kai and Liu, Yi-Ming},
  journal        = {Sensors},
  title          = {Single-Pixel Imaging with Origami Pattern Construction},
  year           = {2019},
  issn           = {1424-8220},
  number         = {23},
  volume         = {19},
  article-number = {5135},
  doi            = {10.3390/s19235135},
  url            = {https://www.mdpi.com/1424-8220/19/23/5135},
}

@Article{HadRussDoll,
  author  = {Sun MJ; Meng LT; Edgar MP; et al},
  journal = {Scientific Reports},
  title   = {A Russian Dolls ordering of the Hadamard basis for compressive single-pixel imaging},
  year    = {2017},
  month   = {06},
  number  = {3464},
  volume  = {7}
}

@Article{reviewsinglepixel,
  author  = {Gibson G M; Johnson S D; Padgett M J},
  journal = {Opt. Exp.},
  title   = {Single-pixel imaging 12 years on: a review},
  year    = {2020},
  number  = {28190},
  volume  = {28}
}

@article{Walsh,
author = {Beer,Tom },
title = {Walsh transforms},
journal = {American Journal of Physics},
volume = {49},
number = {5},
pages = {466-472},
year = {1981},
doi = {10.1119/1.12714},

URL = { https://doi.org/10.1119/1.12714},
eprint = {         https://doi.org/10.1119/1.12714}
}

@Article{CakeCutting,
  author         = {Yu, Wen-Kai},
  journal        = {Sensors},
  title          = {Super Sub-Nyquist Single-Pixel Imaging by Means of Cake-Cutting Hadamard Basis Sort},
  year           = {2019},
  issn           = {1424-8220},
  number         = {19},
  volume         = {19},
  article-number = {4122},
  doi            = {10.3390/s19194122},
  url            = {https://www.mdpi.com/1424-8220/19/19/4122},
}

@Article{ota,
  author  = {Ota ,K ; Hayasaki, Y},
  journal = {Opt. Lett.},
  title   = {Complex amplitude single-pixel imaging},
  year    = {2018},
  pages   = {3682-3685},
  volume  = {43} 
}

@Book{databook,
  author    = {Brunton S; Kutz J},
  publisher = {Cambridge: Cambridge University Press},
  title     = {Data-driven Science and Engineering: Machine Learning, Dynamical Systems and Control},
  year      = {2019}
}

@misc{spgl1site,
   author = {E. van den Berg and M. P. Friedlander},
   title = {{SPGL1}: A solver for large-scale sparse reconstruction},
   note = {https://friedlander.io/spgl1},
   month = {12},
   year = {2019}
}

@article{BergFriedlander:2008,
  Author = {E. van den Berg and M. P. Friedlander},
  Title = {Probing the Pareto frontier for basis pursuit solutions},
  year = {2008},
  journal = {SIAM Journal on Scientific Computing},
  volume = {31},
  number = {2},
  pages = {890-912},
  url = {http://link.aip.org/link/?SCE/31/890},
  doi = {10.1137/080714488}
}

@article{Yang_L1mini,
author = {Yang, Allen and Zhou, Zihan and Balasubramanian, Arvind and Sastry, Shankar and Ma, Yi},
year = {2013},
month = {05},
pages = {},
title = {Fast L1-Minimization Algorithms For Robust Face Recognition},
volume = {22},
journal = {IEEE transactions on image processing : a publication of the IEEE Signal Processing Society},
doi = {10.1109/TIP.2013.2262292}
}

@article{Ota:18,
author = {Kazuki Ota and Yoshio Hayasaki},
journal = {Opt. Lett.},
number = {15},
pages = {3682--3685},
publisher = {OSA},
title = {Complex-amplitude single-pixel imaging},
volume = {43},
month = {08},
year = {2018},
url = {http://www.osapublishing.org/ol/abstract.cfm?URI=ol-43-15-3682},
doi = {10.1364/OL.43.003682},
}

@article{YAO2019483,
title = {DR2-Net: Deep Residual Reconstruction Network for image compressive sensing},
journal = {Neurocomputing},
volume = {359},
pages = {483-493},
year = {2019},
issn = {0925-2312},
doi = {https://doi.org/10.1016/j.neucom.2019.05.006},
url = {https://www.sciencedirect.com/science/article/pii/S0925231219306162},
author = {Hantao Yao and Feng Dai and Shiliang Zhang and Yongdong Zhang and Qi Tian and Changsheng Xu}
}

@InProceedings{KingBa15,
  author    = {Kingma, Diederik and Ba, Jimmy},
  booktitle = {International Conference on Learning Representations (ICLR)},
  title     = {Adam: A Method for Stochastic Optimization},
  year      = {2015},
  address   = {San Diega, CA, USA},
  optmonth  = {12},
}

@Techreport{cifar10,
 author = {Krizhevsky, Alex and Hinton, Geoffrey},
 address = {Toronto, Ontario},
 institution = {University of Toronto},
 number = {0},
 publisher = {Technical report, University of Toronto},
 title = {Learning multiple layers of features from tiny images},
 year = {2009},
 title_with_no_special_chars = {Learning multiple layers of features from tiny images}
}

@article{Scheidt:Bessel,
author = {Dennis Scheidt and Alejandro V. Arzola and Pedro A. Quinto-Su},
journal = {Opt. Lett.},
number = {24},
pages = {6360--6363},
publisher = {Optica Publishing Group},
title = {Shaping the angular spectrum of a Bessel beam to enhance light transfer through dynamic strongly scattering media},
volume = {48},
month = {Dec},
year = {2023},
url = {https://opg.optica.org/ol/abstract.cfm?URI=ol-48-24-6360},
doi = {10.1364/OL.502579},
}

@article{Scheidt_2025,
doi = {10.1088/2040-8986/ad9844},
url = {https://doi.org/10.1088/2040-8986/ad9844},
year = {2024},
month = {dec},
publisher = {IOP Publishing},
volume = {27},
number = {1},
pages = {015605},
author = {Scheidt, Dennis and Quinto-Su, Pedro A},
title = {Errors in single pixel laser imaging emerging from spatial size limits in the bucket detector},
journal = {Journal of Optics}
}

@article{Liu2015ASA,
  title={A Simple and Fast Iterative Soft-thresholding Algorithm for Tight Frames in Compressed Sensing Magnetic Resonance Imaging},
  author={Yunsong Liu and Zhifang Zhan and Jian-Feng Cai and Di Guo and Zhong Chen and Xiaobo Qu},
  journal={ArXiv},
  year={2015},
  volume={abs/1504.07786},
  url={https://api.semanticscholar.org/CorpusID:3531282}
}

\end{document}